\documentclass[sigconf,screen]{acmart}
\usepackage{graphicx}
\usepackage{subfiles}
\usepackage{array}
\usepackage{amsmath}
\usepackage{algorithm}
\usepackage{algpseudocode}
\usepackage{multirow}
\usepackage{subfigure}
\usepackage{caption}
\usepackage{flushend}
\usepackage{balance}
\usepackage{diagbox}
\usepackage{ulem}
\usepackage{algorithmicx}
\usepackage{colortbl}
 \usepackage{enumitem}
 \usepackage{balance}

\usepackage{fancyhdr}

\usepackage{makecell}

\usepackage{float}

\newcommand{\entire}{173,278}
\newcommand{\samples}{749}

\usepackage{microtype}

\AtBeginDocument{%
  \providecommand\BibTeX{{%
    \normalfont B\kern-0.5em{\scshape i\kern-0.25em b}\kern-0.8em\TeX}}}
    
\setcopyright{acmcopyright}
\acmPrice{15.00}
\acmDOI{10.1145/3468264.3468562}
\acmYear{2021}
\copyrightyear{2021}
\acmSubmissionID{fse21main-p221-p}
\acmISBN{978-1-4503-8562-6/21/08}
\acmConference[ESEC/FSE '21]{Proceedings of the 29th ACM Joint European Software Engineering Conference and Symposium on the Foundations of Software Engineering}{August 23--28, 2021}{Athens, Greece}
\acmBooktitle{Proceedings of the 29th ACM Joint European Software Engineering Conference and Symposium on the Foundations of Software Engineering (ESEC/FSE '21), August 23--28, 2021, Athens, Greece}

\begin{document}

\title{A First Look at Developers’ Live Chat on Gitter}



\author{Lin Shi}
\email{shilin@iscas.ac.cn}
\affiliation{Institute of Software Chinese Academy of Sciences, University of Chinese Academy of Sciences
\country{China}
}

\author{Xiao Chen}
\email{chenxiao2021@iscas.ac.cn}
\affiliation{Institute of Software Chinese Academy of Sciences, University of Chinese Academy of Sciences
\country{China}
}

\author{Ye Yang}
\email{yyang4@stevens.edu}
\affiliation{School of Systems and Enterprises, Stevens Institute of Technology
\city{Hoboken}
\state{NJ}
\country{USA}}

\author{Hanzhi Jiang}
\author{Ziyou Jiang}
\email{{hanzhi2021,ziyou2019}@iscas.ac.cn}
\affiliation{Institute of Software Chinese Academy of Sciences, University of Chinese Academy of Sciences
\country{China}
}

\author{Nan Niu}
\email{nan.niu@uc.edu}
\affiliation{Department of EECS, University of Cincinnati
\city{Cincinnati}
\state{OH}\\
\country{USA}}

\author{Qing Wang}
\email{wq@iscas.ac.cn}
\affiliation{State Key Laboratory of Computer Science, Institute of Software Chinese Academy of Sciences, University of Chinese Academy of Sciences
\country{China}
}

\authornote{Corresponding author.\\ }

\renewcommand{\shortauthors}{Lin Shi , Xiao Chen, Ye Yang, Hanzhi Jiang, Ziyou Jiang, Nan Niu, and Qing Wang}
\begin{abstract}
Modern communication platforms such as Gitter and Slack play an increasingly critical role in supporting software teamwork, especially in open source development.
Conversations on such platforms often contain intensive, valuable information that may be used for better understanding OSS developer communication and collaboration. 
However, little work has been done in this regard. 
To bridge the gap, this paper reports a first comprehensive empirical study on developers' live chat, investigating when they interact, what community structures look like, which topics are discussed, and how they interact. We manually analyze 749 dialogs in the first phase, followed by an automated analysis of over 173K dialogs in the second phase. We find that developers tend to converse more often on weekdays, especially on Wednesdays and Thursdays (UTC), that there are three common community structures observed, that developers tend to discuss topics such as API usages and errors, and that six dialog interaction patterns are identified in the live chat communities. 
Based on the findings, we provide recommendations for individual developers and OSS communities,
highlight desired features for platform vendors, and shed light on future research directions. We believe that the findings and insights will enable a better understanding of developers' live chat, pave the way for other researchers, as well as a better utilization and mining of knowledge embedded in the massive chat history.
\end{abstract}


\maketitle

\section{Introduction}
\begin{figure}[hbtp]
\centering
\includegraphics[width=\columnwidth,height=4cm]{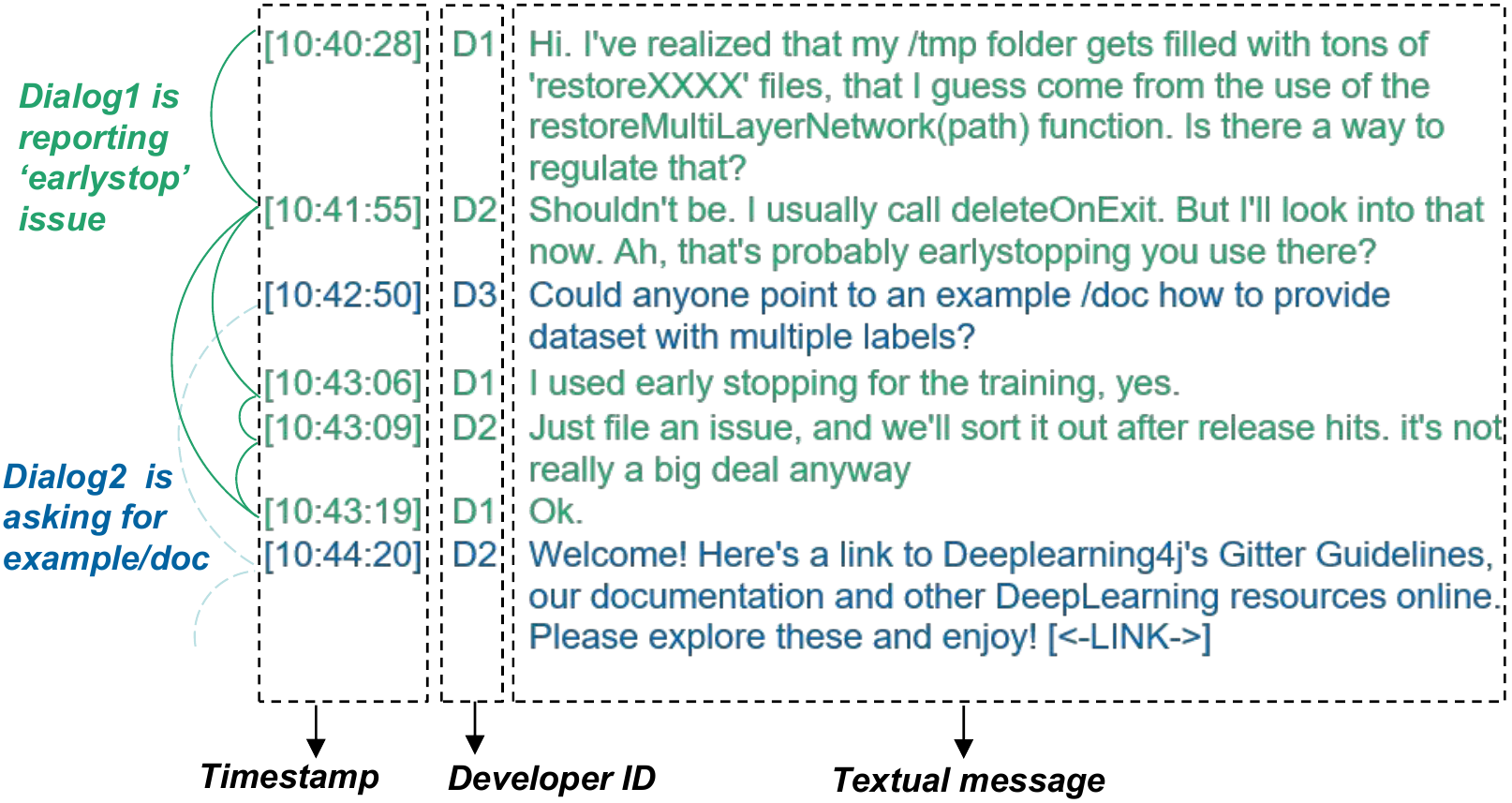}
\vspace{-0.5cm}
\caption{A slice of live chat log from the DL4J community.}
\label{fig:eg}
\vspace{-0.5cm}
\end{figure}

More than ever, online communication platforms, such as Gitter, Slack, Microsoft Teams, Google Hangout, and Freenode, play a fundamental role in team communications and collaboration.
As one type of synchronous textual communication among a community of developers, live chat allows developers to receive real-time responses from others,
replacing asynchronous communication like emails in some cases \cite{DBLP:conf/cscw/LinZSS16,DBLP:conf/msr/ShihabJH09,DBLP:conf/icsm/ShihabJH09}. This is especially true for open source projects that are contributed by globally distributed developers, as well as for many companies allowing developers to work from home due to the COVID-19 pandemic.
Conversations from online communication platforms contain rich information for studying developer behaviors. Figure \ref{fig:eg} exemplifies a slice of live chat log from the \textit{Deeplearning4j} Gitter community. Each utterance consists of a timestamp, developer ID, and a textual message. In addition, two dialogs are embedded in the chat log. The first one is reporting an issue about `earlystop', and the second one is asking for documentation 
\vspace{-0.3cm}
\vspace{0.3cm}
support.
Thus, valuable information, such as when OSS developers interact, what are community structures, which topics are discussed, and how OSS developers interact, can be derived from the massive live chat data, which are important for learning knowledge from productive and effective communication styles, improving existing live chat platforms, and guiding research directions on promoting efficient and effective OSS collaboration.  

Despite that a few empirical studies started to advocate the usefulness of the conversations for understanding developer behaviors \cite{DBLP:conf/icsm/ShihabJH09,DBLP:conf/otm/YuRMM11,DBLP:conf/cscw/LinZSS16}, little focuses on how and what the developers communicate in live chat. 
The most related research is reported by Shihab \textit{et al.} \cite{DBLP:conf/msr/ShihabJH09} on Internet Relay Chat to analyze content, participants, and styles of communications. However, their subjects are IRC meeting logs, which are different in many aspects from developer live chat conversations. Another thread of related work by Parra \textit{et al.} \cite{DBLP:conf/msr/ParraEH20} and Chatterjee \textit{et al.} \cite{DBLP:conf/msr/ChatterjeeDKP20} presents two datasets of open source developer communications in Gitter and Slack respectively, with the purpose of highlighting that live developer communications are untapped information resources. This motivates our study to derive a deeper understanding about the nature of developer communications in open-source software. 

In this paper, we conduct a first comprehensive empirical study on developers' live chat on Gitter, investigating four characteristics: when they interact (communication profile), what community structures look like (community structure), which topics are discussed (discussion topic), and how they interact (interaction pattern).
To that end, we first collect a large scale of developer daily chat from eight popular communities. Then we manually disentangle 749 dialogs, and select the best disentanglement model from four state-of-the-art models according to their evaluation results on the 749 dialogs. After automatic disentanglement, we perform an empirical study on live chat aiming to reveal four characteristics: communication profile, community structure, dialog topic, and interaction pattern.
In total, we studied 173,278 dialogs, 1,402,894 utterances, contributed by 95,416 users from eight open source communities. The main results include: 
(1) Developers are more likely to chat on workdays than weekends, especially on Wednesday and Thursday (UTC); 
(2) Three social patterns are observed in the OSS community of live chat: Polaris network, Constellation network, and Galaxy network;
(3) The top three topics that developers frequently discuss in live chat are API usages, errors, and background information;
and (4) Six interaction patterns are identified in live chat including exploring solution, clarifying answer, clarifying question, direct/discussed answer, self-answered monologue, and unanswered monologue.
The major contributions of this paper are as follows.
\begin{itemize}
    \item We conduct a first large scale analysis study on developers' live chat messages, providing {empirically-based 
    quantitative and qualitative results towards better understanding developer communication profiles, community structures, discussion topics, and interaction patterns.}
    \item We provide practical insights on productive dialogs for individual developers and OSS communities, highlight desired features for platform vendors, and shed light on future directions for researchers.
    \item We provide a large-scale dataset\footnote{https://github.com/LiveChat2021/LiveChat\#5-download} of live chat to facilitate the replication of our study and future applications.
\end{itemize}

In the remainder of the paper, Section II illustrates the background. Section III presents the study design. Section IV describes the results and analysis. Section V is the discussion of results and threats to validity. Section VI introduces the related work. Section VII concludes our work.

\section{Background}
This section describes related key concepts and technologies.
\begin{figure*}[t]
\centering
\includegraphics[width=0.9\textwidth,height=3.9cm]{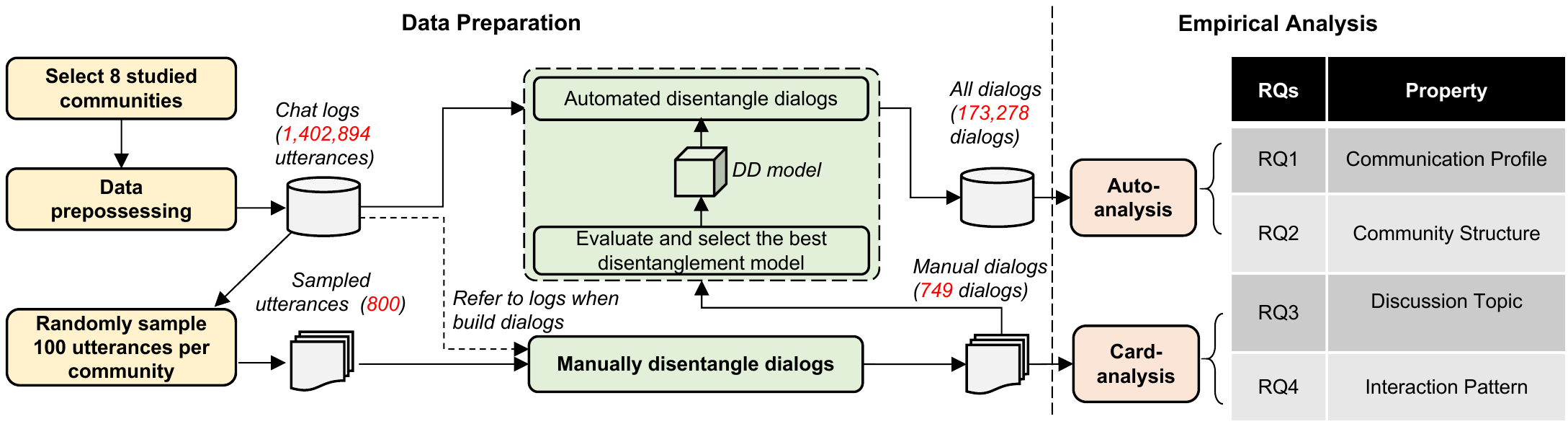}
\vspace{-0.3cm}
\caption{Overview of research methodology}
\label{fig:overview}
\vspace{-0.3cm}
\end{figure*}






 \subsection{The Gitter Platform}
Many OSS communities utilize Gitter \cite{gitter.im} or Slack \cite{slack.com} as their live communication means. In particular, Gitter is currently the most popular online communication platform \cite{DBLP:conf/icse/KaferGB0R18} since it provides open access to public chat rooms and free access to historical data \cite{DBLP:conf/msr/ParraEH20}. 
Considering the popular, open, and free access nature, we conduct this study based on {Gitter}\footnote{In Slack, communities are controlled by the team administrators, whereas in Gitter, access to the chat data is public.}. 
Communities in Gitter usually have multiple chatting rooms: one general room and several specific-topic rooms. Typically, the general room contains most of the participants.
In this study, we only focus on the general rooms.
In total, there are 2,171 communities in Gitter that can be publicly accessed.
The total number of participants of the 2,171 communities is 733,535 as of Nov. 20, 2020.

Three main concepts about Gitter live chat log are concerned in the scope of this study, including \textit{chat log, utterance, and dialog}. In Gitter, developer conversations in one chatting room are recorded in a chat log. As illustrated in Figure \ref{fig:eg}, a typical live chat log contains a sequential set of utterances in chronological order.  
Each \textit{utterance} consists of a timestamp, developer id, and a textual message initiating a question or responding to an earlier message. A \textit{chat log} typically contains a large number of utterances, and at any given time, multiple consecutive utterances might be possibly responding to different threads of dialog discussions. 
The interleaving nature of utterances leads to entangled \textit{dialogs}, as illustrated with the two colors in Figure \ref{fig:eg}. The two colors are used to highlight utterances belonging to two different dialogs, where the links between two utterances indicate the responding relationship.

 \subsection{Challenges in Chat Analysis}
Different from many other sources of software development related communication, the information on online communication platforms are shared in an unstructured, informal, and interleaved manner. Thus, analyzing live chat is quite challenging due to the following barriers.
 \textbf{(1) Entangled dialogs}. 
 Utterances in chat logs form stream information, with dialogs often entangling such as a single conversation is interleaved with other dialogs, as shown in Figure \ref{fig:eg}. It is difficult to perform any kind of high-level dialog analysis without dividing utterances into a set of distinct dialogs. 
\textbf{(2) Expensive human effort.} 
Chat logs are typically high-volume and contain informal dialogs covering a wide range of technical and complex topics. Analyzing these dialogs requires experienced analysts to spend a large amount of time so that they can understand the dialogs thoroughly. Thus, it is very expensive to conduct a comprehensive study on developers' live chat. 
\textbf{(3) Noisy data}.
There exist noisy utterances such as duplicate and unreadable messages in chat logs that do not provide any valuable information.
The noisy data poses a difficulty to analyze and interpret the communicative dialogs.
Next, we will introduce several existing techniques that can automatically disentanglement dialogs.

    
\vspace{-0.3cm}
\subsection{Dialog Disentanglement (DD)}


Four state-of-the-art techniques have been proposed to address the entangled dialog challenge in the natural language processing area:
(1) \textbf{BiLSTM model} \cite{10.1007/978-981-10-6520-0_17} 
predicts whether there exists an edge between two utterances, where the edge means one utterance is a response to another. It employs a bidirectional recurrent neural network with 160 context maximum size, 200 neurons with one hidden layer. The input is a sequence of 512-dimensional word vectors; 
(2) \textbf{BERT model} \cite{DBLP:conf/naacl/DevlinCLT19} 
predicts the probability of utterance $u_i$'s clustering label under the context utterances $u_{\leq i}$ and labels $y_{<i}$. It utilizes the \textit{Masked Language Model} and \textit{Next Sentence Prediction} \cite{DBLP:conf/naacl/DevlinCLT19} to encode the input utterances, with 512 embedding size and 256 hidden units; 
(3) \textbf{E2E model} \cite{DBLP:conf/ijcai/LiuSGLWZ20} performs the dialog Session-State encoder to predict dialog clusters, with 512 embedding size, 256 hidden neurons and 0.05 noise ratio; 
and (4) \textbf{FF model} \cite{acl19disentangle} is a feedforward neural network with two layers, 256-dimensional hidden vectors, and softsign non-linearities. The input is a 77-dimensional numerical feature extracted from the utterance texts, which includes TF-IDF, user name, time interval, whether two utterances contain the same words, \textit{etc.}
In addition, there are four clustering
metrics that are widely used for DD evaluation: Normalized Mutual Information (NMI) \cite{DBLP:journals/jmlr/StrehlG02}, Adjusted Rand Index (ARI) \cite{DBLP:conf/icann/SantosE09}, Shen-F value \cite{DBLP:conf/sigir/ShenYSC06} and F1 score \cite{Crestani01logicand}. 




\section{Methodology and Study Design}


This study aims to investigate four research questions:

\textbf{RQ1 (Communication Profile)}: \textit{
{Do Gitter communities demonstrate consistent community communication profiles?}}
This research question aims at examining common communication profiles across the eight communities,
particularly the frequent time-frames that the developers are active and the typical time interval of a dialog. 

\textbf{RQ2 (Community Structure)}: \textit{What are the structural characteristics of social networks built from developer live chat data?} To understand community characteristics of live chat networks, we perform social network analysis on live-chat utterances, in which each developer is treated as a node, with edges defined as a pair of developers co-occurring in one or more dialogs.

\textbf{RQ3 (Dialog Topic)}: \textit{What are the primary topic types frequently discussed by developers in live chat?}
This research question is designed to identify {discussion topics} in developers' live chat. There have been studies analyzing discussion topics in open forums \cite{DBLP:conf/icse/AryaWGC19}, emails \cite{DBLP:conf/msr/GuzziBLPD13}, and posts in Stack Overflow \cite{DBLP:journals/ese/HanSWDX20,DBLP:conf/iwpc/BeyerM0P18}. It remains unknown what developers are talking about in live chat. {This study aims at filling the gap and providing a complementary perspective using live-chat as a new data source.}  


\textbf{RQ4 (Interaction Pattern)}: \textit{How do developers typically interact with each other in live chat?}
{This research question intends to uncover underlying interaction patterns which signify how developers typically interact (\textit{e.g.}, initiate discussion, respond to questions and social chat.) with one another throughout a dialog life cycle}. 

\vspace{-0.2cm}
\subsection{Methodology Overview} 
The research methodology follows two phases, as illustrated in Figure \ref{fig:overview}. 
First, in the data preparation phase, a large scale of developer daily chat utterances data are collected from eight active communities, and the raw chat utterances data are processed and transformed into associated dialogs using two approaches, \textit{i.e.}, manual screening on a randomly sampled small dataset and automated analysis employing the identified best DD model on the whole dataset. 
Second, in the empirical analysis phase, further analysis will be designed and conducted on these two datasets in order to investigate the characteristics and develop a better understanding of developer live chat data with respect to the research questions.



\newcommand{\tabincell}[2]{\begin{tabular}{@{}#1@{}}#2\end{tabular}}
\begin{table}[bp]
\caption{The statistics of the studied Gitter communities} 
\vspace{-0.3cm}
\label{tab:proj}
 \resizebox{\columnwidth}{!}{
\begin{tabular}{|c|c|c|c|c|c|c|c|}
\hline
\multirow{2}{*}{\textbf{Community}} & \multirow{2}{*}{\textbf{Domain}} & \multicolumn{3}{c|}{\textbf{\tabincell{c}{Entire\\ Population}}} & \multicolumn{3}{c|}{\textbf{\tabincell{c}{Sample\\ Population}}} \\
\cline{3-8}
& &\textbf{P} & \textbf{D} & \textbf{U}& \textbf{P} & \textbf{D} & \textbf{U} \\ \hline
Angular\cite{angular}          & \tabincell{c}{Frontend\\Framework}        & 22,467                 & 79,619            & 695,183              & 125                       & 97                       & 778                         \\ \hline
Appium\cite{appium}           & Mobile                     & 3,979                  & 4,906             & 29,039               & 73                                      & 87                       & 724                         \\ \hline
DL4J\cite{deeplearning4j}             & Data Science               & 8,310                  & 27,256            & 252,846              & 93                       & 100                       & 1,130                         \\ \hline
Docker\cite{docker}           & DevOps                     & 8,810                  & 3,954             & 22,367               & 74               & 90                       & 1,126                         \\ \hline
Ethereum\cite{ethereum}         & \tabincell{c}{Blockchain\\Platform} & 16,154                 & 17,298            & 91,028               & 116                       & 96                       & 516                         \\ \hline
Gitter\cite{gitter}           & \tabincell{c}{Collabration\\Platform}      & 9,260                  & 7,452             & 34,147               & 87                       & 86                       & 515                         \\ \hline
Nodejs\cite{nodejs}           & \tabincell{c}{Web Application\\Framework}  & 18,118                 & 13,981            & 81,771               & 144                       & 98                       & 737                         \\ \hline
Typescript\cite{typescript}       & \tabincell{c}{Programming\\Language }                  & 8,318                  & 18,812            & 196,513              & 110              & 95                       & 1,700                         \\ \hline
\multicolumn{2}{|c|}{\textit{Total}}          & \textit{95,416}        & \textit{173,278}  & \textit{1,402,894}    & \textit{822}             & \textit{749}             & \textit{7,226}               \\ \hline
\end{tabular}
}
\vspace{-0.3cm}
\end{table}

\subsection{Data Preparation}

\textbf{Studied communities.}
To identify studied communities, we select the Top-1 most participated communities from eight active domains, covering front-end framework, mobile, data science, DevOps, blockchain platform, collaboration, web app, and programming language. 
Then, we collect the daily chat utterances from these communities. Gitter provides REST API \cite{gitter.rest} to get data about chatting rooms and post utterances. In this study, we use the REST API to acquire the chat utterances of the eight selected communities, and the retrieved dataset contains all utterances as of ``2020-11-20''. 
Detailed statistics are shown in Table \ref{tab:proj}, where P refers to the number of participants, D refers to the number of dialogs, and U refers to the number of utterances. The total number of participants for the eight communities is 95,416, accounting for 13\% of the total population in Gitter. Thus, we consider that the eight communities are representative of the Gitter platform.

\textbf{Prepossessing the textual utterances.} We first normalize non-ASCII characters like emojis to standard ASCII strings. Some low-frequency tokens contribute little to the analysis of live chat, such as URL, email address, code, HTML tags, and version numbers. We replace them with specific tokens \textit{<URL>, <EMAIL>, <HTML>, <CODE>, and <ID>}. We utilize Spacy \cite{spacy.io} to tokenize sentences into terms and perform lemmatization and lowercasing on terms with Spacy to alleviate the influence of word morphology.



\textbf{Manual labeling of dialog disentanglement.}
We employ a 3-step manual process to generate a sample dialog dataset for further analysis.
First, we randomly sample 100 utterances from each community's live chat log, with the intention to trace corresponding dialogs associated with each of the 100 utterances. This step leads to a total of 800 utterances from the entire 1,402,894 utterances of the eight communities. 
Next, using each utterance as a seed, we identify its {preceding and successive} utterances iteratively so that we can group related utterances into the same dialog as complete as possible. Specifically, for each utterance, we 
determine its context by examining the consecutive chats in the chat log, and manually link it to related utterances belonging to the same dialog. 
Then, the next step is cleaning. Specifically, we excluded unreadable dialogs: (1) dialogs that are written in non-English languages; (2) dialogs that contain too much code or stack traces; (3) Low-quality dialogs such as dialogs with many typos and grammatical errors; and (4) Dialogs that involve channel robots. 
After these steps, we include additional six thousand utterances which are associated with the initial 800 utterances. This leads to a total of 7,226 utterances, manually disentangled into 749 dialogs, as summarized in Table 1. 
Note that, removing bot-involved dialogs has little impact on our results. First, the bot-involved dialogs are relatively small in volume. In our study, only one of the eight projects utilizes bots, and more specifically, only nine out of 800 sampled dialogs are excluded due to bot involvement. Second, we observed that the bot-generated utterances are rather trivial, such as greeting information, links to general guidelines, and status updates.

To ensure the correctness of the disentanglement results, a labeling team was put together, consisting of one senior researcher and six Ph.D. students. All of them are fluent in English, and have done either intensive research work with software development or have been actively contributing to open-source projects. The senior researcher trained the six Ph.D. candidates on how to disentangle dialogs and provided consultation during the process. The disentanglement results from the Ph.D. candidates were reviewed by others. We only accepted and included dialogs to our dataset when the dialogs received full agreement. When a dialog received different disentanglement results, we hosted a discussion with all team members to decide through voting. The average Cohen’s Kappa about dialog disentanglement is 0.81.

\textbf{Automated Dialog Disentanglement.} {To analyze the dialogs on a large scale, we {experiment with the four state-of-art DD approaches, as introduced in Section 2.2 (\textit{i.e.} BiLSTM model, Bert model, E2E model, and FF model). Specifically, we use the manual disentanglement sample data from the previous step as ground truth data, compare and select the best DD model for the purpose of further analysis in this study.} 
The comparison results from our experiments show that the \textbf{FF approach} significantly outperforms the others on disentangling developer live chat by achieving the highest scores on all the metrics. The average scores of NMI, Shen-F, F1, and ARI are 0.74, 0.81, 0.47, and 0.57 respectively\footnote{Due to space, experimental details on evaluation existing DD models are provided on Github: https://tinyurl.com/3dyu5n44}. 
Finally, we use the best FF model to disentangle all the 1,402,894 utterances in chat logs. In total, we obtain 173,278 dialogs.} 
\vspace{-0.2cm}
\subsection{Empirical Analysis Design}
\subsubsection{Analysis for RQ1 (Communication Profile)} 
{As good communication habits suggest more productive development practices,} we intend to reveal the temporal communication profiles of developers, including when the developers are active and how long the respondent replies to the dialog initiator.
First, we collect all the utterance time of the entire population, and analyze the peak hours and peak days. Then we calculate the response time lag :
\begin{equation}
    \begin{aligned}
       & Time\_lag=T_r-T_i
    \end{aligned}
\end{equation}

\noindent
where $T_i$ is the time that the initiator launched the dialog, and $T_r$ is the time the first respondent replied. 
We automatically calculate the utterance times and response time lags for all the {\entire} dialogs. 

\subsubsection{Analysis for RQ2 (Community Structure)}
We aim to visualize the social networks of developers in live chat, and summarize the common structures. Social network analysis (SNA) describes relationships among social entities, as well as the structures and implications of their connections \cite{DBLP:books/cu/WF1994}. 
For studying relationships among developers in one OSS community, we generate the social networks according to the following definition:
\begin{equation}
\begin{aligned}
   & G=\{V,E\} \\
   & V=\{d_1,d_2,...,d_n\}\\
   & E=\{<d_j,d_k>\} 
\end{aligned}    
\end{equation}

\noindent
where $d_i$ is a developer in the chatting room, $d_j$ is one dialog initiator, and $d_k$ is a respondent to $d_j$. 
Specifically, for each disentangled dialog, we first identify its initiator and all the respondents. The initiator is the developer who launches the dialog, and the respondents are other developers who participate in the dialog. 
Then we add a link between the initiator and each respondent. 
(Note that, RQ2 focuses on exploring responding behaviors, i.e., the interaction between initiators and respondents, thus the links/edges are defined only between these two roles. Additionally, we will explore the interaction relationship among initiators and all responders in RQ4, which focuses on discussion behaviors.)
Finally, we employ an unweighted graph when constructing the social networks, for visualizing the relationship of all the developers in live chat. The social network could exhibit the connectivity and density of the open-source community.
We build the social networks based on all the 173,278 dialogs {by using the automatic graph tool, Gephi \cite{bastian2009gephi}}.

To understand the topology of the eight social networks, we report the following SNA measures that have been widely used by previous studies \cite{DBLP:journals/ijseke/SchreiberZ20, DBLP:conf/sigsoft/MeneelyWSO08}. 
Note that, we excluded developers who never received replies (AKA. Haircut) when calculating SNA measures following previous work \cite{bird2006mining,sulistianingsih975performance}.
(1) \textbf{Degree} \cite{diestel2005graph}
measures the number of edges. 
(2) \textbf{Betweenness} \cite{freeman1977set} measures the frequency that a developer lies on the shortest path between other developers.
(3) \textbf{Closeness} \cite{bavelas1950communication} 
{measures the average farness (inverse distance) of one developer to all other developers.} 
(4) \textbf{Diameter} \cite{bouttier2003geodesic} is the largest geodesic distance in the connected network.
(5) \textbf{Clustering coefficient} \cite{watts1998collective} is a measure of the degree to which developers in a graph tend to cluster together.

\subsubsection{Analysis for RQ3 (Discussion Topic)} 

Our goal is identifying {discussion topics} in developers' live chat. Chatterjee \textit{et al.} \cite{DBLP:conf/msr/ChatterjeeDKP20} observed that live chats provide similar information as can be found in Q\&A posts on Stack Overflow. 
{To effectively identify and organize dialog topics}, we extend Beyer \textit{et.al}'s category of question categories on Stack Overflow \cite{DBLP:conf/iwpc/BeyerM0P18}. 
We choose Beyer \textit{et.al}'s category for two reasons.
First, since developers use both Question and Answer forums (such as Stack Overflow) and live chat to resolve development issues, we consider the categories of Stack Overflow questions are partially applicable to the dialog topics in live chat. 
Second, Beyer \textit{et.al}'s category harmonizes five taxonomies presented in previous studies \cite{DBLP:conf/msr/AllamanisS13,Beyer17,DBLP:conf/icsm/BeyerP14,DBLP:journals/ese/RosenS16,DBLP:conf/icse/TreudeBS11} that are already validated and suitable to the posts of developers' questions. 

\begin{figure*}[tp]
\centering
\subfigure[Hourly distribution]{
        \label{fig:hours}
        \includegraphics[width=0.67\columnwidth,height=4cm]{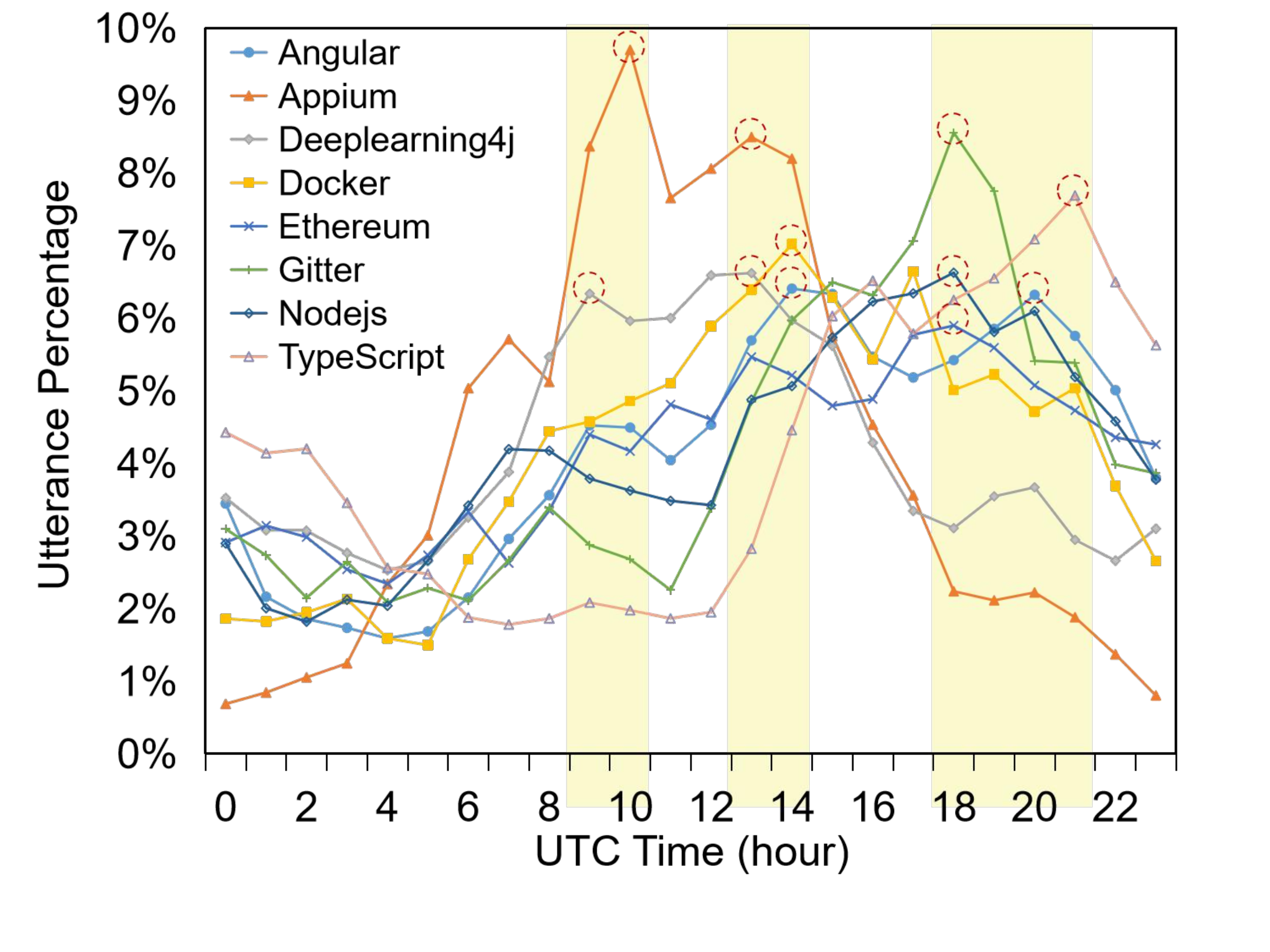}
    
}
\subfigure[Day of week distribution]{
        \label{fig:days}
        \includegraphics[width=0.67\columnwidth,height=4cm]{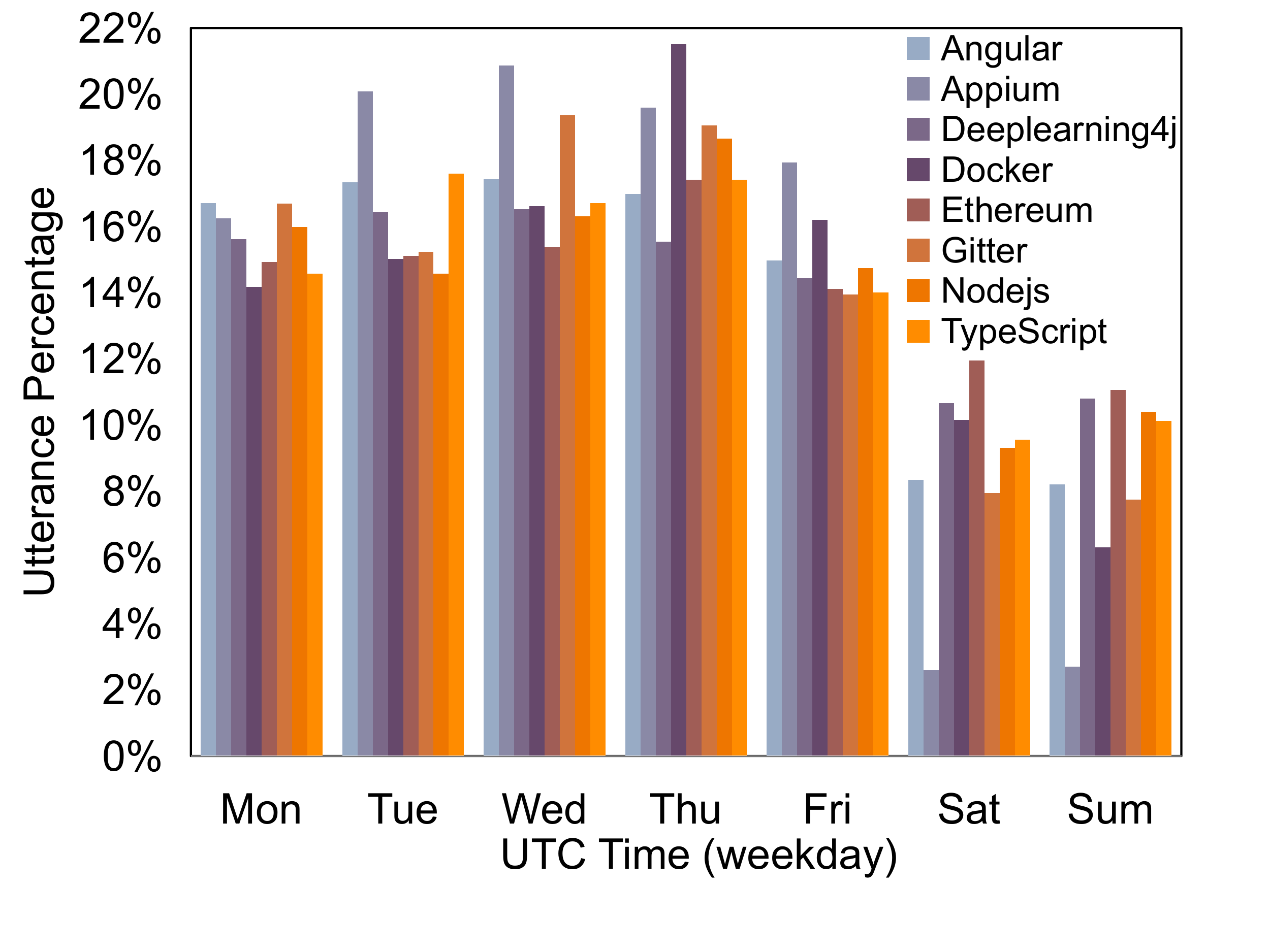}
    
}
\subfigure[Frequency distribution of response times]{
        \label{fig:timelag}
        \includegraphics[width=0.67\columnwidth,height=4cm]{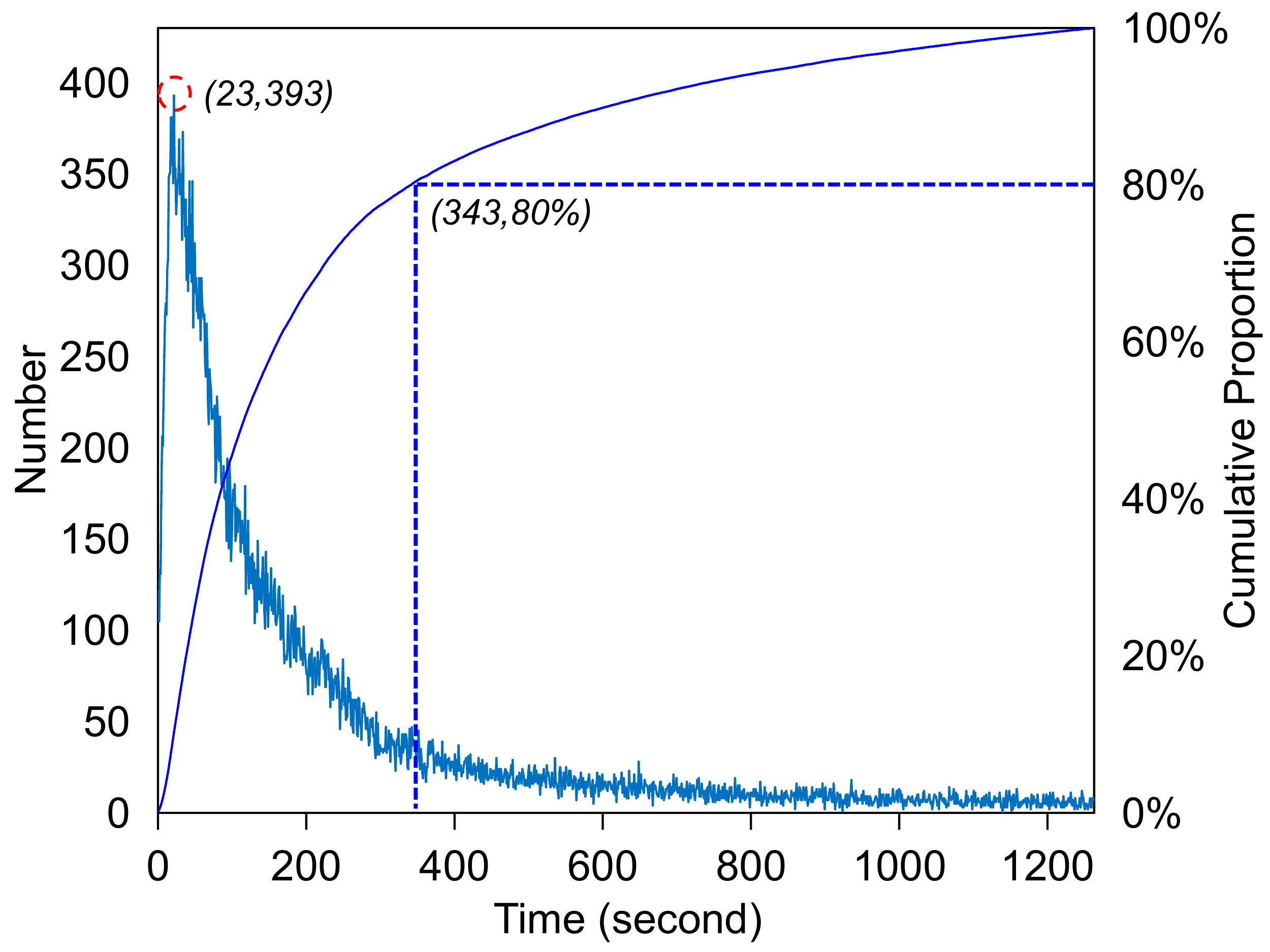}
}
\vspace{-0.5cm}
\caption{Statistic results about communication profiles}
\label{fig:RQ2}
\vspace{-0.5cm}
\end{figure*}

Nonetheless, the predefined category is not meant to be comprehensive,  
thus, we employ a \textit{hybrid card sort} process \cite{DBLP:journals/es/FincherT05} to manually determine the topics of dialogs. In a \textit{hybrid card sort}, the sorting begins with the predefined Beyer \textit{et.al}'s category and participants could create their own as well. The newly-created topic is instantly updated into the topic set and can be used by other participants then.
The participants are the same team that manually disentangle dialogs, and the labeling process is similar to manual dialog disentanglement as introduced in Section 3.2. 
{Specifically, the sorting process is conducted in one round, with a concluding discussion session to resolve the disagreement in labels based on majority voting. The average Cohen’s Kappa about dialog topics is 0.86.}


\begin{table}[bp]
\caption{Developer intent category in live chat}
\vspace{-0.3cm}
\label{tab:intent}
\resizebox{\columnwidth}{!}{
\begin{tabular}{@{}cll@{}}
\toprule
\textbf{Code} & \multicolumn{1}{c}{\textbf{Label}} & \multicolumn{1}{c}{\textbf{Description}}                     \\ \midrule
OQ            & Original Question                 & The first question from the developer to initiate the dialog \\
CQ            & Clarifying Question                & Developers ask for clarifications                            \\
FD            & Further Details                    & Developers provide more details                              \\
FQ            & Follow Up Question                 & Developers ask for follow up questions about relevant issues \\
PA            & Potential Answer                   & A potential answer or solution provided by developers        \\
PF            & Positive Feedback                  & Developer provides positive feedback for working solutions  \\
NF            & Negative Feedback                  & Developer provides negative feedback for useless solutions   \\
GG            & Greetings/Gratitude                & Greetings or expressing gratitude                            \\ \bottomrule
\end{tabular}
}
\end{table}

\subsubsection{Analysis for RQ4 (Interactive Pattern)} 
Live-chat conversations generally serve the purposes of solution exploration and discussion stimulation. To uncover underlying patterns that shape and/or direct more productive conversations, we first adopt a developer intent codebook \cite{DBLP:conf/chiir/Qu0CZTQ19} and manually label the interaction links that appeared in each dialog. The developer intent codebook is built from previous work on user intent in information-seeking conversations \cite{DBLP:conf/chiir/Qu0CZTQ19}, as summarized in Table \ref{tab:intent}.


Then, we employ an \textit{open card sort} \cite{DBLP:journals/es/RuggM05} process to assign an interactive pattern to a dialog based on the sequence of developers' intents.
In an open sort, the sorting begins
with no predefined patterns and participants develop their own
patterns. The two participants individually assigned patterns to the same dialogs. The sorting process is conducted in multiple rounds. 
In the first round, all participants label dialogs of one community, with an intensive discussion session to achieve conceptual coherence about patterns. A shared pool of patterns is utilized and carefully maintained, and each participant could select existing patterns from and/or add new pattern names into the shared pool. 
Then we divide into two teams to label the remaining dialogs. 
Each dialog will receive two pattern labels, and we resolve disagreement based on majority voting. The average Cohen’s Kappa about interactive patterns is 0.82.


After identifying underlying interaction patterns, we further explore their statistical characteristics in aspects of distribution and duration. We calculate the  duration of a dialog as follows:
\vspace{-5pt}
\begin{equation}
    Duration = T_e-T_i
\end{equation}
where $T_e$ is the time that the dialog ended, and $T_i$ is the time that the initiator launched the dialog. This metric can reflect the life cycle of one dialog.

Note that, to keep the workload of manually labeling each dialog manageable, we answer RQ3 and RQ4 by manually analyzing the {\samples} sampled dialogs.
We believe that although we could only manually analyze a small percentage of the disentangled dialogs, this dataset supports our methodology as being useful for discovering valuable findings.

\section{Results and Analysis}
\subsection{RQ1: Communication Profile} 
{To answer this question, we analyze two metrics, \textit{i.e.} utterance time and response time. Next, we report the results of comparing these metrics across the eight Gitter communities.}

\textbf{Utterance Time.} Figure \ref{fig:hours} compares the distribution of utterances' intensity over 24 hours, across the eight communities. {First, we identify the peak hours of each community in red dashed circles, then highlight the time windows based on the peak hours contained in it with the yellow shade.} We can see that, there are three windows of peak hours, which are from UTC 9 to 10, 13 to 14, and 18 to 21. In addition, UTC 1 to 6 corresponds to the low chatting-activity hours. Developers are less active in chatting at that time. 
Figure \ref{fig:days} shows the distribution of the utterances across different weekdays. We can see that developers chat more on workdays than on weekends (UTC).

\textbf{Response Time.} 
Figure \ref{fig:timelag} exhibits the distribution of response time calculated from the {\entire} dialogs of the eight communities.
The average response time is 220 seconds, the maximum time lag is 1,264 seconds, and the minimum time lag is two seconds. 
The peak point is (23, 393), which means there are 393 dialogs that got replies in 23 seconds. 
We can see that, the time lag largely increases from 0 to 23 seconds, and descend in a long tail.
Eighty percent of the dialogs get first responses in 343 seconds. 
As reported by a recent study on Stack Overflow \cite{DBLP:conf/icsm/0003MZ0WL20}, the threshold of fast answers was 439 seconds. In comparison, live chat gets 50\% faster ((439-220)/439) replies than the fast answers in Stack Overflow. Therefore, we consider the responses from the live chat are relatively fast.


\textbf{Answering RQ1:} The peak hours for live chat are from UTC 9 to 10, 13 to 14, and 18 to 21, while UTC 1 to 6 is the low-active hours. 
Developers are more likely to chat on workdays than weekends, especially on Wednesdays and Thursdays (UTC). Moreover, live chat gets 50\% faster replies than the fast answers in Stack Overflow.
\begin{table*}[bhtp]
\caption{Social Network measures of the eight communities}
\label{tab:RQ1}
 \resizebox{0.6\textwidth}{!}{
\begin{tabular}{|c|c|c|c|c|c|c|c|c|}
\hline
                                                                           & \multicolumn{4}{c|}{Constellation}                                                                                                    & \multicolumn{3}{c|}{Polaris}                                                                        & Galaxy                          \\ \cline{2-9} 
\multirow{-2}{*}{}                                                         & \textbf{Angular}                & \textbf{DL4J}                   & \textbf{Nodejs}                 & \textbf{Typescript}             & \textbf{Appium}                 & \textbf{Docker}                 & \textbf{Gitter}                 & \textbf{Ethereum}               \\ \hline
\textit{Init. \%}                                                          & \cellcolor[HTML]{FEF2EA}54.54\% & \cellcolor[HTML]{FFFFFF}51.38\%    & \cellcolor[HTML]{F4AF81}70.06\% & \cellcolor[HTML]{FCE8DA}56.94\% & \cellcolor[HTML]{F19A5F}75.04\% & \cellcolor[HTML]{F19557}76.22\% & \cellcolor[HTML]{F1995D}75.33\% & \cellcolor[HTML]{ED7D31}81.70\% \\ \hline
\textit{Resp. \%}                                                          & \cellcolor[HTML]{ED7D31}13.90\% & \cellcolor[HTML]{F3A570}11.39\% & \cellcolor[HTML]{FDECE1}6.83\%  & \cellcolor[HTML]{F19556}12.42\% & \cellcolor[HTML]{FCE3D2}7.43\%  & \cellcolor[HTML]{FFFCF9}5.84\%  & \cellcolor[HTML]{F8CAAA}9.04\%  & \cellcolor[HTML]{FFFFFF}5.59\%     \\ \hline
\textit{Both \%}                                                           & \cellcolor[HTML]{F29C61}31.56\% & \cellcolor[HTML]{ED7D31}37.23\% & \cellcolor[HTML]{F8C8A8}23.11\% & \cellcolor[HTML]{F2A069}30.65\% & \cellcolor[HTML]{FCE6D7}17.53\% & \cellcolor[HTML]{FCE4D4}17.94\% & \cellcolor[HTML]{F6BB93}25.62\% & \cellcolor[HTML]{FFFFFF}12.71\%    \\ \hline
\textit{Degree}                                                            & \cellcolor[HTML]{ED7D31}9.15     & \cellcolor[HTML]{F8C8A7}5.02     & \cellcolor[HTML]{FDF1E8}2.75     & \cellcolor[HTML]{FAD7BF}4.2      & \cellcolor[HTML]{FEF3EC}2.59     & \cellcolor[HTML]{FFFFFE}1.96     & \cellcolor[HTML]{FFFDFB}2.07     & \cellcolor[HTML]{FFFFFF}1.92     \\ \hline
\textit{Betweenness}                                                       & \cellcolor[HTML]{FFFBF8}0.000273 & \cellcolor[HTML]{F5B58A}0.000867 & \cellcolor[HTML]{FCE4D4}0.000468 & \cellcolor[HTML]{F7BF99}0.000786 & \cellcolor[HTML]{F2A16A}0.001035 & \cellcolor[HTML]{F09151}0.001173 & \cellcolor[HTML]{ED7D31}0.001342 & \cellcolor[HTML]{FFFFFF}0.000231 \\ 
\hline
\textit{Closeness}                                                         & \cellcolor[HTML]{FAD4BB}0.35     & \cellcolor[HTML]{EF8843}0.42     & \cellcolor[HTML]{FFFFFF}0.31     & \cellcolor[HTML]{FBDFCC}0.34     & \cellcolor[HTML]{FAD4BB}0.35     & \cellcolor[HTML]{F6BE98}0.37     & \cellcolor[HTML]{ED7D31}0.43     & \cellcolor[HTML]{FBDFCC}0.34     \\ \hline
\textit{Diameter}                                                          & \cellcolor[HTML]{FBE3D2}8       & \cellcolor[HTML]{FFFFFF}6          & \cellcolor[HTML]{F19A5F}13      & \cellcolor[HTML]{F9D4BB}9       & \cellcolor[HTML]{F7C6A4}10      & \cellcolor[HTML]{F7C6A4}10      & \cellcolor[HTML]{FBE3D2}8       & \cellcolor[HTML]{ED7D31}15      \\ \hline
\textit{\begin{tabular}[c]{@{}c@{}}Clustering \\ coefficient\end{tabular}} & \cellcolor[HTML]{F5B285}0.41     & \cellcolor[HTML]{ED7D31}0.6      & \cellcolor[HTML]{FFFDFB}0.14     & \cellcolor[HTML]{F8C8A8}0.33     & \cellcolor[HTML]{FDECE1}0.2      & \cellcolor[HTML]{FFFFFF}0.13     & \cellcolor[HTML]{FDECE1}0.2      & \cellcolor[HTML]{FEF7F2}0.16     \\ \hline
\end{tabular}
}
\end{table*}

\vspace{-0.7cm}
\subsection{RQ2: Community Structure}
To answer RQ2, we first examine the structural properties of the developer social networks across the eight communities, and then try to draw some common observations based on these social networks.



\textbf{Properties of social networks.}
Table \ref{tab:RQ1} shows the social network properties of the eight communities. Init.\%, Resp.\%, and Both\% denote the percentage of developers serving the role of dialog initiators, respondents, and both.
Intuitively, we consider that respondents share their knowledge with others, while initiators receive knowledge from others. We can see that, the four communities (Appium, Docker, Gitter, and Ethereum) have a higher percentage (75.04\%-81.70\%) of dialog initiators and a lower percentage (18.30\%-24.96\%) of respondents/both. 
The high percentage of dialog initiators may relate to the applicable nature of the open-source projects, \textit{e.g.}, Ethereum is one of the most widely used open-source blockchain systems, thus there are a large number of users acquiring technical support from live chat.
The other four communities (Angular, DL4J, Nodejs, and Typescript) have a higher percentage (29.94\%-48.62\%) of respondents/both. A possible explanation is that these four projects are more widely used for development purposes, \textit{e.g.}, Angular is a platform for building mobile and desktop web applications, therefore, such communities appear to be knowledge-sharing and collaborative.
\begin{figure*}[tp]
\centering
\subfigure[Constellation \{{from left to right:} Angular, DL4J, Nodejs, Typescript \}]{
        \includegraphics[width=2\columnwidth]{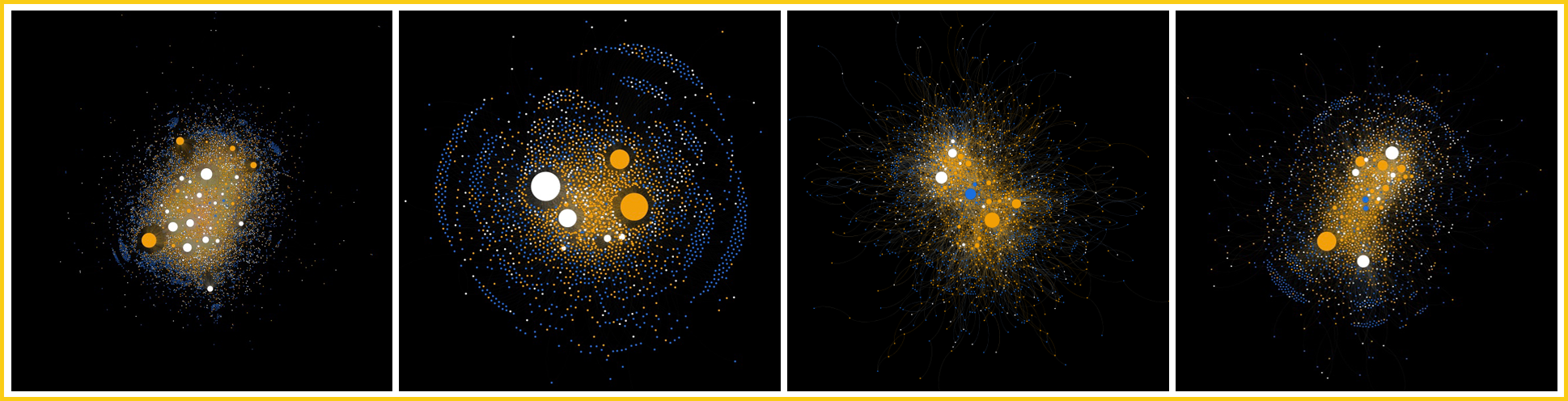}
    
}
\subfigure[Polaris \{{from left to right:} Appium, Docker, Gitter \} ]{
        \includegraphics[width=1.47\columnwidth]{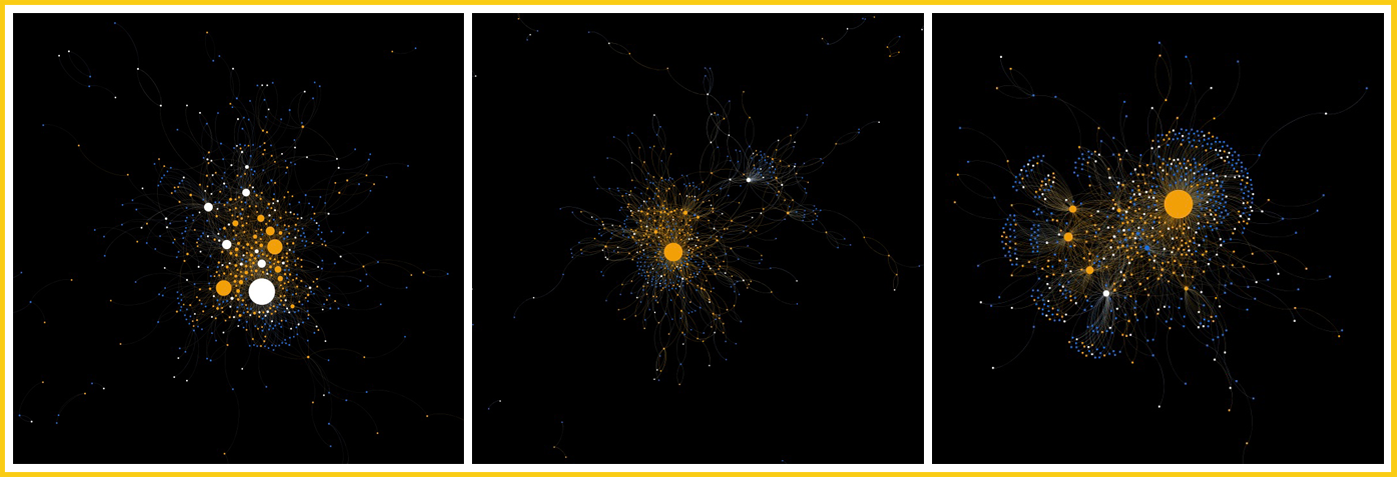}
    
}
\subfigure[Galaxy \{ Ethereum \}]{
        \includegraphics[width=0.502\columnwidth]{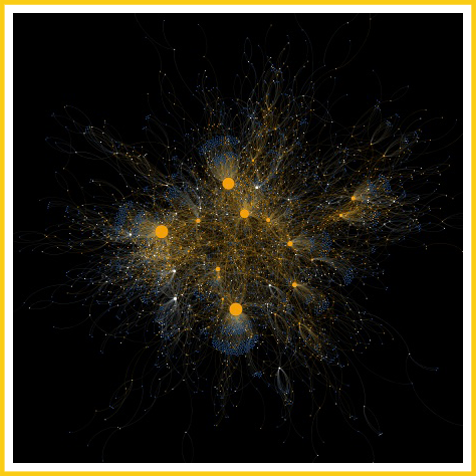}
}
\vspace{-0.4cm}
\caption{{Visualization of the eight developer live-chat social networks}}
\label{fig:network}
\vspace{-0.4cm}
\end{figure*}


\textbf{{Categorizing developer} social networks.} Figure \ref{fig:network} shows the social network visualizations of the eight communities generated by Gephi. Each node represents one developer, and the edge denotes the dialog relationship between two developers. 
We color the vertex of the initiator with blue, the vertex of the respondent with white, and the vertex of both roles with orange. In addition, the node’s size indicates its corresponding \textit{degree}. 
Based on the observation on community structures, we categorize the eight communities into three groups, consisting of: 
(1) \textbf{Polaris network} is a type of highly centralized network where the community is organized around its single focal point;  
(2) \textbf{Constellation network} is a type of moderately centralized network where the community is organized around its multiple focal points;
and (3) \textbf{Galaxy network} is a type of decentralized network where all individuals in the community have similar relationships.
In Figure \ref{fig:network}, the four communities on the top (\textit{i.e.}, Angular, DL4J, NodeJS, and Typescript) belong to the Constellation network, \textit{i.e.}, moderately centralized network. Three communities
(\textit{i.e.}, Appium, Docker, and Gitter) belong to the Polaris network, \textit{i.e.}, a highly centralized network. The remaining Ethereum community belongs to the Galaxy network, \textit{i.e.}, decentralized network.  
Previous studies have shown that a highly centralized network may reflect an uneven distribution of knowledge across the community, where knowledge is mostly concentrated at the focal points \cite{Manju1998Network,Krebs04}.
Therefore, the three Polaris communities (Appium, Docker, and Gitter) may have a higher risk of single-point failure, if the focal developer is inactive, whereas the Galaxy network (Ethereum) has the lowest risk, followed by the Constellation network (Angular, DL4J, NodeJS, and Typescript). 


In Table \ref{tab:RQ1}, we can also see that the Constellation networks and Polaris networks have higher scores in terms of average degree (1.96-9.15), betweenness (0.000273-0.001342), and closeness (0.31-0.43). The phenomena indicate that the focal points in Constellation networks and Polaris networks make the communities more connected. A study on email-connected social networks \cite{bird2006mining} shows that the mean betweenness of developers is 0.0114, which on average is higher than live chat communities. 
{Nodes with high betweenness may have considerable influence within a network in allowing information to pass from one part of the network to the other. Lower betweenness indicates that developers in the live chat may have less influence than developers in email in spreading information.} 
However, the average in-degree and out-degree of networks built on emails are significantly lower, with 0.00794 and 0.00666 for developers. {While developers in live chat have more concentration and higher density, along with the closeness centrality values, indicating a more closely connected community than that from email.}
Developers in Constellation communities have higher clustering coefficient scores (0.14-0.60), {indicating that developers in Constellation communities are more densely connected, \textit{i.e.}, the developers of Angular, DL4J, Nodejs, and Typescript know each other better than the others.}

\textbf{Answering RQ2:} By visualizing the social networks of eight studied communities, we identify three social network structures for developers' live chat. Half of the communities (4/8) are Constellation networks. A minority of the communities (3/8) are Polaris networks. Only one community belongs to the Galaxy network. In comparison, we find that developers in the live chat may have less influence than developers in email in spreading information, but have a more closely connected community than that from email. 
\subsection{RQ3: Discussion Topic}
\begin{figure}[t]
\centering
\includegraphics[width=1\columnwidth]{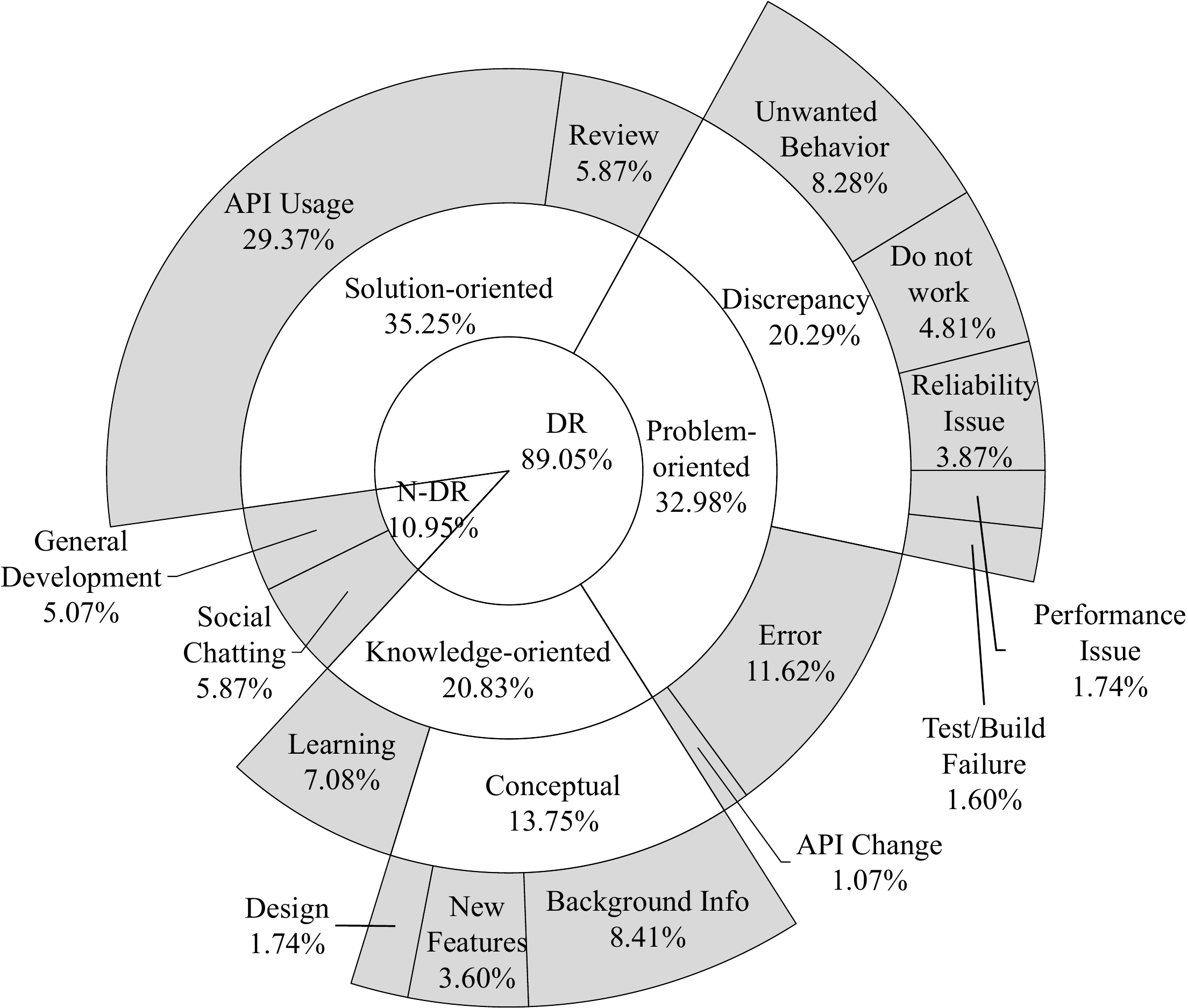}
\vspace{-3ex}
\setlength{\belowcaptionskip}{-10pt}
\caption{Distribution of discussion topics in developer live chat {by reading from center to outside}}
\label{fig:donut}
\end{figure}

Figure \ref{fig:donut} shows the distribution of discussion topics in developer live chat.
The figure shows discussion topics in gray and their categories in white, as well as the percentages of the corresponding dialogs. The taxonomy expands outwards from higher-level categories to lower-level categories and topics. 
{In this study, we extend the Beyer \textit{et al.}'s category to accommodate the open
and live chat by: 
(1) adding social chatting and general development categories;
and (2) decomposing ``Conceptual'' and ``Discrepancy'' categories to distinguish more valuable information such as unwanted behavior and new features.}
For more information on the dialog topics, we provide a public Github repository with details and examples\footnote{https://github.com/LiveChat2021/LiveChat\#34-rq3-discussion-topic}.

The most inner circle shows that,  across all eight communities, 89.05\% of dialogs are domain-related (DR) topics such as topics related to the business domain of the community, while 10.95\% of dialogs are non-domain related (N-DR) topics such as general development or social chatting.
{DR topics can be further decomposed into} three sub-categories 
based on their different purposes. These include solution-oriented dialogs which have the highest proportion (35.25\%), followed by problem-oriented dialogs  (32.98\%) and knowledge-oriented dialogs (20.83\%). 
Among the 35.25\% \textit{solution-oriented} dialogs, 29.37\% are about \textit{API usage}, and 5.87\% are about \textit{Review}.
Among the 32.98\% \textit{problem-oriented} dialogs, most of them (20.29\%) discuss discrepancy, consisting of \textit{unwanted behavior, do not work, reliability issue, performance issue}, and \textit{test/build failure}. We can see that, developers discuss more `unwanted behavior' and `do not work', than reliability issues, performance issues, and test/build failures.
Among the 20.83\% knowledge-oriented dialogs, most of them (13.75\%) discuss conceptual, consisting of \textit{background info, new features}, and \textit{design}. 
Overall, the top three frequent topics are \textit{API usage (29.37\%), Error (11.62\%)}, and \textit{Background info (8.41\%)}.


\textbf{Answering RQ3:} Developers launch solution-oriented dialogs and problem-oriented dialogs more than knowledge-oriented dialogs. Nearly 1/3 of dialogs are about API usage. Developers discuss more error, unwanted behavior, and do-not-work, than reliability issues, performance issues, and test/build failures.

\subsection{RQ4: Interaction Pattern}

\begin{figure}[b]
\centering
\includegraphics[width=\columnwidth]{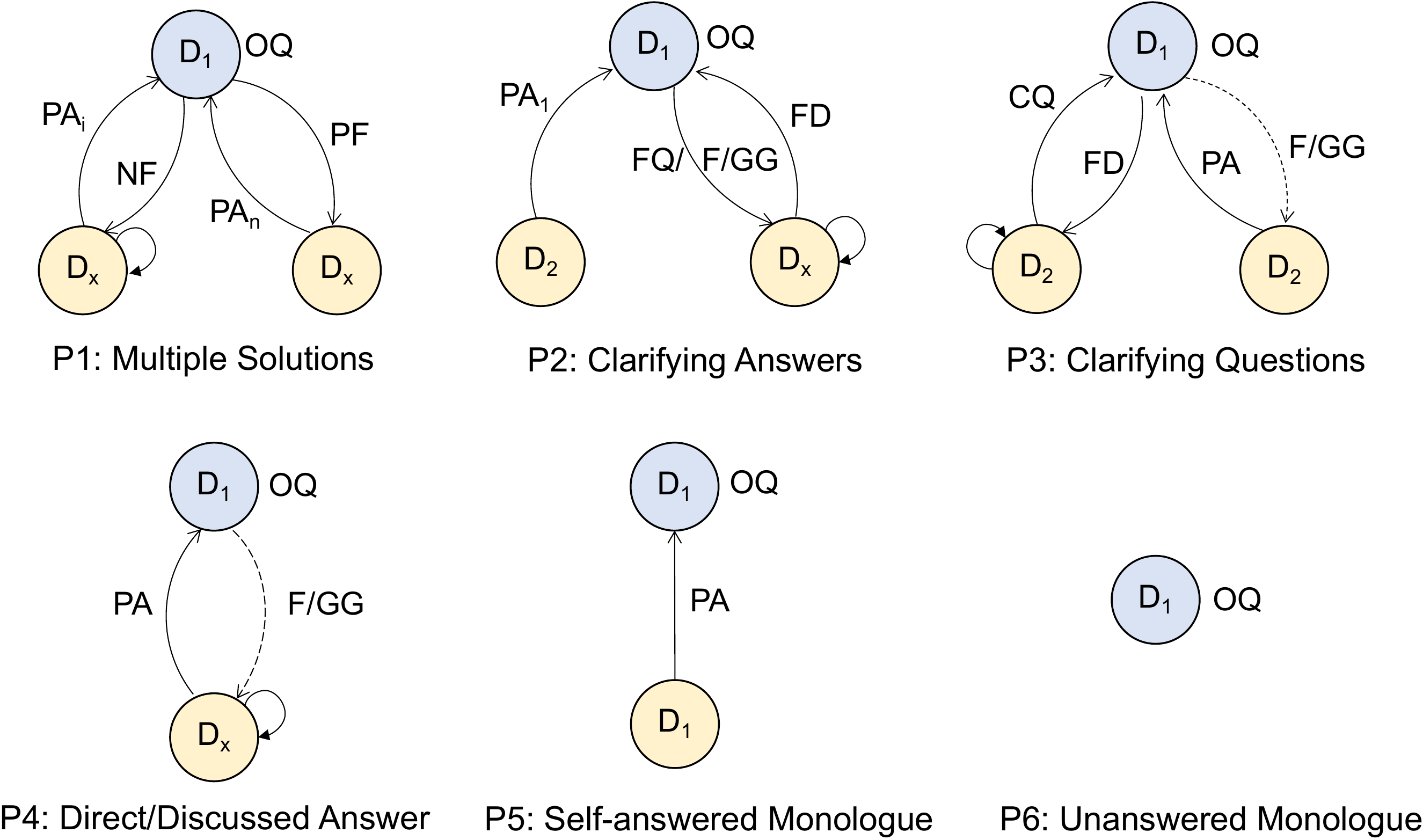}
\vspace{-0.5cm}
\caption{Interactive patterns, F denotes feedback including negative feedback and positive feedback, dashed lines denote optional interaction.}
\label{fig:pattern}
\vspace{-0.3cm}
\end{figure}

\begin{figure}[t]
\centering
\includegraphics[width=\columnwidth]{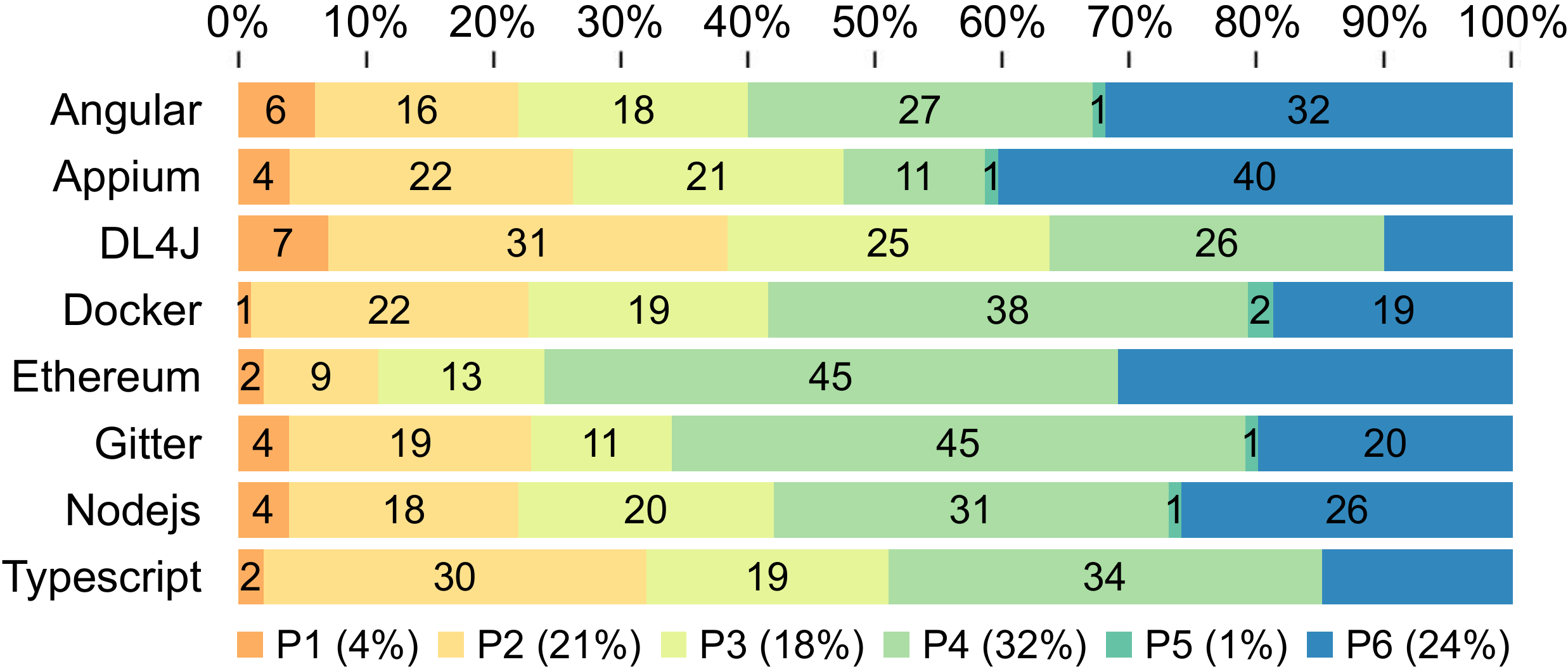}
\vspace{-3ex}
\caption{Distribution of interaction patterns among different communities}
\label{fig:topic-pattern}
\vspace{-0.4cm}
\end{figure}

\begin{figure}[b]
\centering
\includegraphics[width=0.9\columnwidth]{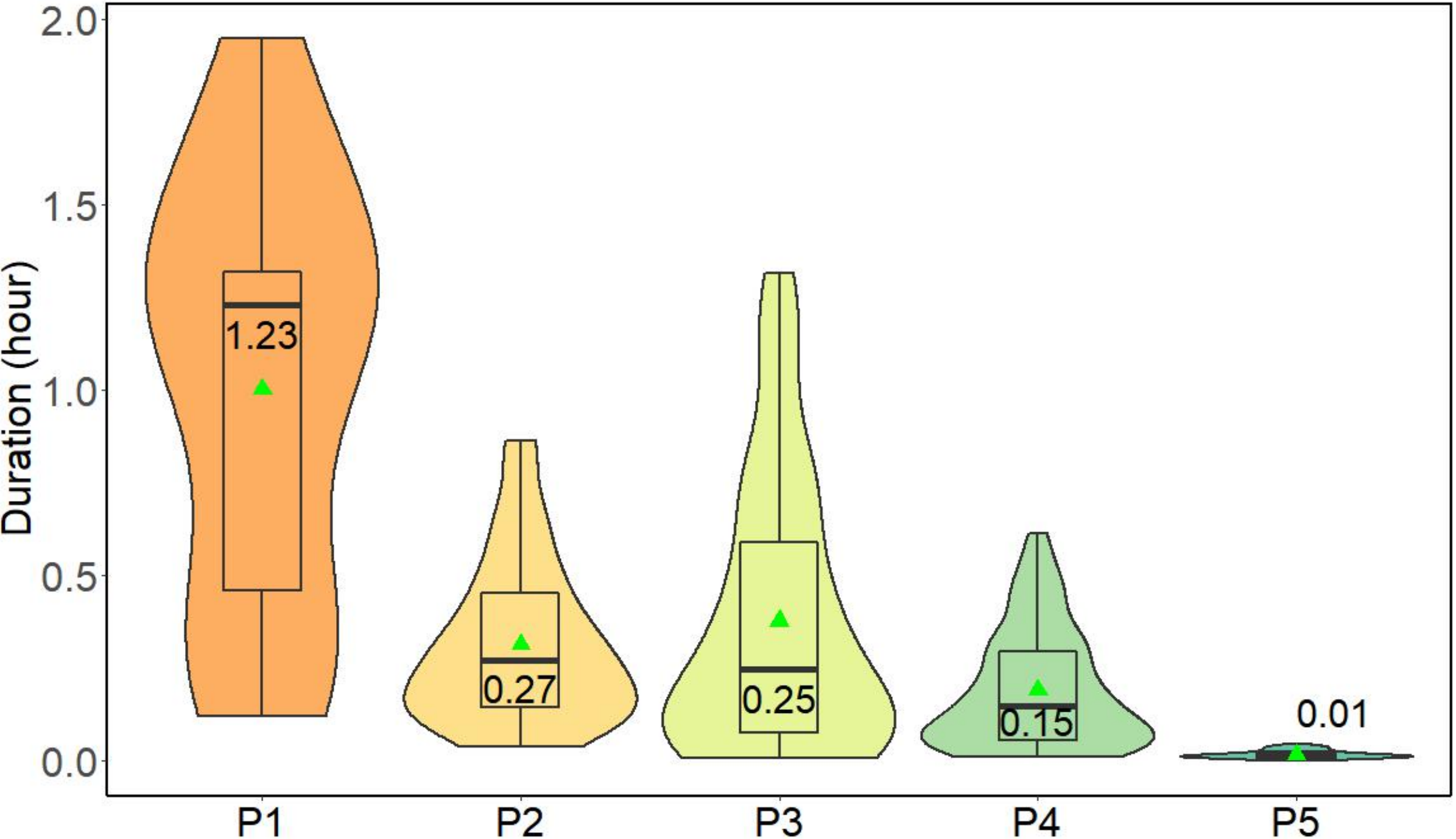}
\vspace{-0.2cm}
\caption{Distribution of duration for interaction patterns, except for unanswered monologues.}
\label{fig:duration}
\vspace{-0.2cm}
\end{figure}

\textbf{Interaction patterns.} Figure \ref{fig:pattern} illustrates the six interaction patterns in live chat, constructed using open card sorting as introduced in Section 3.3.4. This figure shows dialog initiators in blue nodes, respondents in yellow nodes. The lines denote the reply-to relationships, and the labels represent developer intents in Table \ref{tab:intent}. In this work, we identify the following six interaction patterns: 
(1) \textbf{P1: Exploring Solutions}. Given the original questions posted by the dialog initiator, other developers provide possible answers. But the initiator gives negative feedback indicating these answers do not address the question. When the correct answer is posted, the initiator gives positive feedback and ends the dialog. 
(2) \textbf{P2: Clarifying Answer}. Given the original questions posted by the dialog initiator, another developer provides a possible answer. Then the initiator posts follow-up questions to clarify the answer until the initiator fully understands. 
(3) \textbf{P3: Clarifying Question.} Given the original questions posted by the dialog initiator, the respondent requires the initiator to clarify the question in more detail until they fully understand. Then the respondent posts an answer, and the initiator gives feedback or greetings.
(4) \textbf{P4: Direct/Discussed Answer.} Given the original questions posted by the dialog initiator, the respondent directly answers, or answers after an internal discussion.  
(5) \textbf{P5: Self-answered Monologue.} The original questions posted by the dialog initiator are answered by himself or herself. 
(6) \textbf{P6: Unanswered Monologue.} The original questions posted by the dialog initiator are not answered.

\textbf{Percentage of patterns.} Figure \ref{fig:topic-pattern} shows the percentage of interaction patterns in different communities, and the average percentages are shown in the legends. 
P1\textasciitilde P6 refer to the six interaction patterns defined above.
We can see that the \textit{direct/discussed answer} (P4) pattern takes the largest proportions in most communities. 
In addition, we note that quite a few dialogs (1\%) belong to self-answered monologue, while 24\% of dialogs belong to unanswered monologue. Nearly 1/4 of dialogs did not get responses in live chat. We will discuss more monologue in Section 5.1.

\textbf{Duration of patterns.} Figure \ref{fig:duration} shows the violin plots with the distribution of duration for each pattern.
P1\textasciitilde P5 refer to interaction patterns defined above. Here we only exhibit five patterns because the P6 refers to unanswered monologues which barely have a duration.
We can see that although P1 takes a small proportion in dialogs, it lasts the longest. Its average duration is 1.00 hours. 
P2 and P3 last slightly longer than P4. P5 lasts the shortest, and its average duration is 0.02 hour.

\textbf{Answering RQ4:} Six interaction patterns are identified in live chat: \textit{exploring solutions, clarifying answer, clarifying question, direct/discussed answer}, \textit{self-answered monologue}, and \textit{unanswered monologue}. The \textit{direct/discussed answer} pattern takes the largest proportions in most communities. There are still 1/4 dialogs that did not get responses on average. Dialogs that belong to the \textit{Exploring Solutions} pattern last the longest time than others.


\section{Discussion}
In this work, we take a first look at developers’ live chat on Gitter in terms of communication profile, community structure, discussion topic, and interaction pattern. Our work paves the way for other researchers to be able to utilize the same methods in other software communities. Additionally, as communication is a large part of successful software development, and Gitter is one of the main platforms for communication of GitHub users, it is important to explore how software engineers use Gitter and their pain points of using it. Aiming at promoting efficient and effective OSS communications, we discuss the main implications of our findings for OSS developers, communities, platform vendors, and researchers. 


\subsection{Individual Developers} 
Based on our findings, we present the following implications for individual OSS developers to attract attentions and receive responses effectively and efficiently.

\textbf{(1) Provide example code or data when seeking solution help (RQ3, RQ4).} In Figure \ref{fig:donut}, we reported that, there are nearly 1/3 dialogs are problem-oriented. In the live chat, it is important to provide example code in problem-oriented dialogs, to make other developers quickly understand and avoid missing key information. This finding is also in line with the evidence provided by previous studies \cite{DBLP:journals/infsof/CalefatoLN18,6624013} on Stack Overflow. 
Here is an example of showing 
 \begin{center}
\fbox{\parbox{0.95\columnwidth}{ 
\small
\noindent
\ttfamily{<D1> Hey there, anyone has an idea on this issue: Have quotes showing up in the UI. I tried to remove it by replacing it in the .ts file...I'm not sure what else to try. \\
<D2> I'm having trouble understanding what's going on. Can you toss your example in a plunker?}
}}
\end{center}
the importance of using examples in the live 
chat. The corresponding dialog confirms the Clarifying Question Pattern (P3) with multiple back-and-forth interactions for clarifying questions. If examples are provided at the beginning, the process of issue-resolving would be expedited. Particularly, the Angular community gives great importance to example demonstration. Developers are encouraged to create examples via a demonstration platform, named Plunker \cite{plnkr}.

\textbf{(2) Be aware of low-active hours (RQ1)}. Our results show that developers are more active during some time slices in live chat. Figure \ref{fig:hours} demonstrates the most active time slices are UTC 9-10, 13-14, and 18-21, corresponding to Central European/American daytime or Asia nighttime. 
Noticeably, more developer live chatting happens on Wednesdays and Thursdays than on other weekdays (UTC),  
which possibly corresponds to communication, coordination, and preparation for integration/release deadlines on Fridays.
This observation also confirms the ``commercially viable alternative'' of the OSS projects reported in recent studies \cite{DBLP:conf/icse/ZhangZSWJ20, DBLP:journals/it/Gonzalez-BarahonaR13,DBLP:journals/misq/Fitzgerald06}. One of the common findings is that the traditional notion of OSS projects that are driven by voluntary developers is now outdated. OSS has become a commercially viable alternative, and some OSS projects have become critical building blocks for organizations worldwide.  
For example, Docker is widely used by software companies around the world\footnote{https://www.docker.com/customers}, including Adobe, AT\&T, PayPal, \textit{etc}.
Therefore, instead of chatting on weekends, developers likely discuss their problems in the live chat on workdays.

While low-active time slices (UTC 1-6) mostly correspond to Central European/American nighttime or Asia day time. In cases where developers find issues and need support during low-active hours, we suggest several options. First, it is recommended to simultaneously post questions to other alternative platforms, \textit{e.g.}, issues and emails. Second, they better follow up in live-chat if not receiving timely responses to their questions posted during low-active hours. Finally, employ some automated reminder bot, for example, to review the list of questions posted during low-active hours.


\textbf{(3) Avoid asking amid ongoing discussions (RQ4).} When identifying the unanswered monologues patterns from dialogs, we note that 30\% of them are launched in the middle of ongoing active and intense conversations on a different topic. 
In such cases, new questions are easily flooded by the utterances of the ongoing discussions. Therefore, to increase the opportunity of getting responses, developers could post their questions after the ongoing discussions.
In case that an urgent matter emerges, we suggest that the platform vendors provide special accommodations to flag such urgency and redirect the team’s attention to it, such as multi-threaded conversation (e.g., in Slack) or a highlight tag for urgent questions and let others supervise the usage of an urgent tag to avoid abuse.

\subsection{OSS Communities} 
we provide the following recommendations for OSS managers to improve the management and coordination of the communities.

\textbf{(1) Mitigate the risk of single-point failure (RQ2).}
As reported in RQ2, the three Polaris communities (Appium, Docker, and Gitter) may have a higher risk of single-point failure, if the focal developer left or became inactive. It is noticeable that there are some second focal points smaller than those most focal ones in the three Polaris networks, and this may suggest practical strategies in order to mitigate the risk of single-point failure. For example, the Polaris communities may design and employ appropriate incentives or policies to second focal developers, for improving the resilience of the live chat communities.

\textbf{(2) Improve OSS documentation for newcomers (RQ1, RQ4)}.
It is reported that some newcomers complained that it is hard to start on a new project and get timely help from other community members, which may make them gradually lose motivation, or even give up on contributing \cite{DBLP:conf/sigsoft/TanZS20}. To facilitate newcomers to familiarize themselves and make contributions in a more efficient manner, OSS communities may consider utilizing the results of our study to improve OSS documentation. For example, the results of RQ1 show active and low-active time slices, and the results of RQ4 show many unanswered monologues are asking amid ongoing discussions. That information could be incorporated into the README documents for newcomers, who are looking to contribute to a project and how to get timely help from others. 


\vspace{-1.9ex}
\subsection{Platform Vendors}

This section discusses several desired features for facilitating more productive conversations.
Specifically, these are organized from communication platform vendors' perspective, in support of more intelligent and productive chatting options, leveraging on mining and knowledge sharing of intensive historical conversations.

\textbf{(1) Highlight and organize conversation topics (RQ3).}
As suggested by the previous study, multi-dimensional separation of concerns \cite{DBLP:conf/re/D05} is a powerful concept supporting collaborative development by breaking a large discussion down into many smaller units. It highlights that the online communication platform vendors could provide support for a set of predefined panels that focus on certain topics.
In the results of RQ3, we provide a taxonomy of discussion topics in live chat. 
{The online communication platform vendors could refer to this taxonomy to create topic panels. For example, API-usage panel, Error panel, Background-info panel, \textit{etc}.}
These multiple panels could bring the following benefits to community members: (i) quickly understanding and retrieving the intents of the dialog initiators; (ii) reducing interference and focusing more on topics of interest; and (iii) identifying important information reported by the developers.


\textbf{(2) Annotate important questions (RQ3).} 
When investigating dialog topics, we note that certain types of dialogs suggest information for future software evolution. For example, there are 3.6\% dialogs discussing new features, and 8.28\% discussing unwanted behaviors. These dialogs are valuable for product teams to plan future releases. 
In the meanwhile, other types of dialogs indicate unrevealed defects of existing systems. For example, 11.62\% dialogs discuss errors, 4.81\% discussing something that does not work, 3.87\% discussing reliability issues, and 1.74\% discussing performance issues. Properly annotating those dialogs with ``Feature Request'', ``Enhancement'', and ``Bugs'' would help to preserve valuable information, and contribute to productivity and quality improvement of the software. As an example, techniques for mining live chat have been explored for identifying feature requests from chat logs \cite{DBLP:conf/icse/ShiXLWL020}.

\vspace{-1.9ex}
\subsection{Researchers}
Research in the SE area could dedicate to promote efficient and effective OSS communication in the following directions.


\textbf{(1) Automatically recommend similar questions (RQ4).}
Existing online communication platforms only record the massive history chat messages, but do not consider a deeper utilization of those historical data. Actually, we note that developers post similar questions in live chat sometimes. In DL4J, one initiator posted a question, and he got a reply like this: ``Someone else asked a very similar question a while ago.'' However, it is not easy for the initiator to accurately retrieve the similar question out from the massive history messages. In addition, some questions that got unanswered are largely due to many similar questions being previously answered.
Therefore, we consider that, it will save developers' effort if 
researchers could develop approaches that automatically recommend similar questions and the corresponding discussions. 

\textbf{(2) Automatically assign appropriate respondents (RQ4).} By analyzing the dialogs belonging to exploring solution patterns (P1), we note that respondents who are not quite familiar with the technologies related to the posted questions might give ineffective solutions. Although such discussions could make developers understand the problem better, the multiple fail-and-try interactions still prolong the process of issue-resolving. Therefore, to make conversations more productive, it is expected
to develop approaches that could recommend or assign appropriate respondents according to their historical answers.

\textbf{(3) Automatically push valuable information to project repositories (RQ3).}
Valuable information such as feature requests or issue reports, either manually annotated by developers or automatically detected by tools, needs to be well-documented and well-traced in the scope of project repositories. Typically, code repositories such as Github or Gitlab provide the functionality of issue tracking. It would be more efficient if
researchers could provide a convenient way to directly push or integrate the valuable information into the code repository.
In addition, the following linguistic patterns might be helpful to automatically classify dialog topics.
It is observed that questions from the \textit{API usage} category include phrases such as: ``how to do sth?'', ``how can I do sth?'', ``can anyone help me with sth?'', ``is there any way to sth?'', or ``I want/need to do sth''. For example, ``How to bundle my Angular 2 app into a `bundle.js' file?'' and ``Can anyone help me with PWA using angular 2''. 
Questions from the \textit{Error} category are likely to be ``I get/receive this error'', ``Anyone had an error like this'', ``Does anyone know a solution for sth'', or directly posting the specific exception names. For example, ``I receive a NotFound error'' and ``Anyone had an error like this before when trying to load a route?''
Questions from the \textit{Background info} category are likely to be ``what/why/when...'', ``I would like to know sth'', or ``is there sth for...''. For example, ``When Appium will support Xcode 8.2?'' and ``Are there any limitations for automating the iOS app made with Swift 3?''.

\textbf{(4) Analyze effects of social chatting (RQ3).}
As reported in RQ3, 10.95\% of dialogs are non-domain related (N-DR) topics such as general development or social chatting.
A recent study \cite{DBLP:journals/corr/abs-2101-05877} emphasizes an important role of social interactions, such as the simple phrase ``How was your weekend?'', to show peer support for developers working at home during the COVID-19 pandemic.
Future work may explore more patterns and effects of social chatting, \textit{e.g.}, pre/post-pandemic comparison.









\vspace{-0.5ex}
\subsection{Threats To Validity}

\textbf{External Validity}. The external threats relate to the generalizability of the proposed approach. Our empirical study used eight Top-1 most participated open source communities from Gitter. 
Although we generally believe all communities may benefit from knowledge learned from more productive, effective communication styles, future studies are needed to focus on less active communities and comparison across all types of communities.


\textbf{Internal Validity}. The internal threats relate to experimental errors and biases. The first threat relates to the accuracy of the dialog disentanglement model adopted by us. Although we select the best model to disentangle dialogs from the state-of-the-art approaches, the accuracy score for the best model is still not quite satisfactory. It will have an impact on the results of RQ1 and RQ2. To address this issue, one of our ongoing works is to build a new efficient dialog disentanglement model based on deep learning to improve the accuracy of existing disentanglement approaches.
The second threat relates to the random sampling process.
Sampling may lead to incomplete results, \textit{e.g.}, topic taxonomy and interaction patterns. In the future, we plan to enlarge the analyzed dataset and inspect whether new topics or interaction patterns are emerging. 
The third threat might come from the process of manual disentanglement and card sorting. We understand that such a process is subject to introducing mistakes. To reduce that threat, we establish a labeling team, and perform peer-review on each result. We only adopt data that received the full agreement, or reach agreements on different options.


\textbf{Construct Validity}. The construct threats relate to the suitability of evaluation metrics. In this study, manual labeling of topics and interactive patterns is a construct threat. To minimize this threat, we use a well-known approach used by previous work \cite{DBLP:conf/icsm/BeyerP14,DBLP:conf/sigsoft/SafwanS19,DBLP:conf/sigsoft/BagherzadehK19}
to build reasonable taxonomies for textual software artifacts. 
\vspace{-0.2cm}
\section{Related Work}
Our work is related to previous studies that focused on synchronous and asynchronous communication in the OSS community. 

\textbf{Synchronous Communication in OSS community.} 
Recently, more and more work has realized that live chat via modern communication platforms
plays an increasingly important role in team communication. 
Lin \textit{et al.} \cite{DBLP:conf/cscw/LinZSS16} conducted an exploratory study on understanding the role of Slack in supporting software engineering by surveying 104 developers. Their research revealed that developers use Slack for personal, team-wide, and community-wide purposes, and developers use bots for team and task management in their daily lives. They highlighted that live chat plays an increasingly significant role in software development, replacing email in some cases.
Shihab \textit{et al.} \cite{DBLP:conf/icsm/ShihabJH09,DBLP:conf/msr/ShihabJH09} analyzed the usage of developer IRC meeting channels of two large open-source projects from several dimensions: meeting content, meeting participants, their contribution, and meeting styles. Their results showed that IRC meetings are gaining popularity among open source developers, and highlighted the wealth of information that can be obtained from developer chat messages. 
Yu \textit{et al.} \cite{DBLP:conf/otm/YuRMM11} analyzed the usage of two communication mechanisms in global software development projects, which are synchronous (IRC) and asynchronous (mailing list). Their results showed that developers actively use both communication mechanisms in a complementary way.
To sum up, existing empirical analysis of live chat mainly focused on the usage purpose \cite{DBLP:conf/cscw/LinZSS16}, the usage of live meetings \cite{DBLP:conf/icsm/ShihabJH09,DBLP:conf/msr/ShihabJH09}, and comparison with different communication mechanisms and knowledge-share platforms \cite{DBLP:conf/otm/YuRMM11}. 
There is a lack of in-depth analysis of the community properties and the detailed discussion contents. Our study bridges that gap with a large-scale analysis of communication profiles, community structures, dialog topics, and interaction patterns in live chat. 

\textbf{Asynchronous Communication in OSS community.} Prior studies have empirically analyzed asynchronous communication in the OSS community, including mailing-list, issue discussions, and Stack Overflow. 
Bird \textit{et al.} \cite{bird2006mining} mined email social network on the Apache HTTP server project. They reported that the email social network is a typical electronic community: a few members account for the bulk of the messages sent, and the bulk of the replies.
Di Sorbo \textit{et al.} \cite{DBLP:conf/kbse/SorboPVPCG15} proposed a taxonomy of intentions to classify sentences in developer mailing lists into six categories: feature request, opinion asking, problem discovery, solution proposal, information seeking, and information giving. Although the taxonomy has been shown to be effective in analyzing development emails and user feedback from app reviews \cite{panichella2015how}, Huang \textit{et al.} \cite{Huang2018Automating} found that it cannot be generalized to discussions in issue tracking systems, and they addressed the deficiencies of Di Sorbo \textit{et al.}'s taxonomy by proposing a convolution neural network based approach. 
Arya \textit{et al.} \cite{DBLP:conf/icse/AryaWGC19} identified 16 information types, such as new issues and requests, solution usage, \textit{etc.}, through quantitative content analysis of 15 issue discussion threads in Github. They also provided a supervised classification solution by using Random Forest with 14 conversational features to classify sentences.
Allamanis and Sutton \cite{DBLP:conf/msr/AllamanisS13} presented a topic modeling analysis that combines question concepts, types, and code from Stack Overflow to associate programming concepts and identifiers with particular types of questions, such as, ``how to perform encoding''. Similarly, Rosen and Shihab \cite{DBLP:journals/ese/RosenS16} employed Latent Dirichlet Allocation-based topic models to help us summarize the mobile-related questions from Stack Overflow.
Our work differs from existing research in that we focus on synchronous communication which poses different challenges as live chat logs are informal, unstructured, noisy, and interleaved.
%
%
%
%
%
%
%
\section{Conclusion and Future Work}
In this paper, we have presented the first large-scale study to gain an empirical understanding of OSS developers' live chat. Based on 173,278 dialogs taken from eight popular communities on Gitter, we explore
the temporal communication profiles of developers, the social networks and their properties towards the community, the taxonomy of discussion topics, and the interaction patterns in live chat.
Our study reveals a number of interesting findings.
Moreover, we provide recommendations for both OSS developers and communities, highlight advanced features for online communication platform vendors, and provoke insightful future research questions for OSS researchers.
In the future, we plan to investigate how well can we automatically classify the dialogs into different topics, as well as attempt to construct knowledge bases according to already answered questions and their corresponding solutions from live chat.
We hope that the findings and insights that we have uncovered will pave the way for other researches, help drive a more in-depth understanding of OSS development collaboration, and promote a better utilization and mining of knowledge embedded in the massive chat history. 
To facilitate replications or other types of future work, we provide the utterance data and disentangled dialogs used in this study online: \href{https://github.com/LiveChat2021/LiveChat}{https://github.com/LiveChat2021/LiveChat}.

\section*{Acknowledgments}
We deeply appreciate anonymous reviewers for their constructive and insightful suggestions towards improving this manuscript.
This work is supported by the National Key Research and Development Program of China under Grant No. 2018YFB1403400, the National Science Foundation of China under Grant No. 61802374, 62002348, and 62072442, and Youth Innovation Promotion Association CAS.
\bibliographystyle{ACM-Reference-Format}
\bibliography{ref}
\end{document}